\documentclass[sigconf, 10pt, nonacm]{acmart}

\usepackage[english]{babel}
\usepackage{booktabs}
\usepackage{color}
\usepackage[tableposition=below]{caption}
\usepackage{colortbl}
\usepackage{subcaption}
\usepackage{enumitem}
\usepackage{tcolorbox}
\usepackage{multirow}
\usepackage{xcolor,pifont}

\newcommand{\xmark}{\ding{55}}%

\newcommand{\ltgrey}{\rowcolor[gray]{0.88}}

\setcopyright{none}

\renewcommand\footnotetextcopyrightpermission[1]{} 

\copyrightyear{}
\acmYear{}
\setcopyright{none}
\acmConference{}
\acmBooktitle{}
\acmPrice{} 
\acmISBN{}
\acmDOI{}

\definecolor{chicagomaroon}{rgb}{0.5, 0.0, 0.0}
\iftrue
\newcommand{\sumanth}[1]{\textcolor{violet}{\emph{#1 --SR}}}
\newcommand{\alex}[1]{\textcolor{chicagomaroon}{\emph{#1 --AL}}}
\newcommand{\todo}[1]{\textsf{\textcolor{red}{[TODO: #1]}}}
\newcommand{\grant}[1]{\textcolor{teal}{\emph{#1 --GH}}}
\newcommand{\geoff}[1]{\textcolor{purple}{\emph{#1 --GV}}}
\newcommand{\stefan}[1]{\textcolor{green}{\emph{SS: #1}}}
\else
\newcommand{\todo}[1]{}
\newcommand{\sumanth}[1]{}
\newcommand{\grant}[1]{}
\newcommand{\geoff}[1]{}
\newcommand{\stefan}[1]{}
\newcommand{\alex}[1]{}
\fi

\begin{document}


\title{After the Breach: Incident Response within Enterprises}

\author{Sumanth Rao}
\orcid{0000-0002-0075-457X}
\affiliation{%
  \institution{UC San Diego}
  \city{La Jolla}
  \state{CA}
  \country{USA}
}
\email{svrao@ucsd.edu}

\begin{abstract}
  Enterprises are constantly under attack from sophisticated adversaries. These adversaries use a variety of techniques
to first gain access to the enterprise, then spread laterally inside its networks, establish persistence, and finally exfiltrate
sensitive data, or hold it for ransom. While historically, enterprises have used different \textit{Incident Response} systems that monitor hosts, servers, or network devices to detect and report threats, these systems often need many analysts to triage and respond to alerts. However, the immense quantity of alerts to sift through, combined with the potential risk of missing a valid threat makes the task of the analyst challenging. To ease this manual and laborious process, researchers have proposed a variety of systems that perform automated attack investigation. These systems collect data, track causally related events, and present the analyst with an interpretable summary of the attack. In this paper, we present a survey of systems that perform automated attack investigation, and compare them based on their designs, goals, and heuristics. We discuss the challenges faced by these systems, and present a comparison in terms of their ability to address these challenges. We conclude with a discussion of open problems in this space.

\end{abstract}

\maketitle

\section{Introduction}




Attacks that target enterprises are often multi-stage, and involve a variety of creative techniques at each step. Indeed, so are some of the famous hacks in recent years that target software, systems, or accounts. For example, the 2020 SolarWinds hack involved compromising the SolarWinds Orion software, which was used by thousands of organizations. The 2022 Cash App breach involved a former employee accessing financial reports containing user information. The 2023 CitrixBleed hack exploited a vulnerability in Citrix's NetScalar product, and used it to gather sensitive information from several enterprises deployments of the product. While the occurrence of such attacks is nothing new, the scale and sophistication of such attacks has exploded in recent years. \footnote{HHS reports the largest data breaches affected 60\% more individuals in 2023 than in 2022~\cite{hhsransomware:online}.}

Once inside, attackers use a variety of techniques to move laterally through the
network (e.g., pass-the-hash, Kerberoasting, etc), compromise accounts with high
privileges and access to sensitive systems or data, and use those privileges to
ferry data outside the network. Throughout this process, attackers play a
carefully crafted cat-and-mouse game against threat detection systems, trying to
masquerade as legitimate users within the network (e.g., by using legitimate
tools, or probing access to systems infrequently). Indeed, this process can
often be long, and as a result, attackers can persist inside the network for
weeks, if not months before being detected.\footnote{IBM reports the average
time to detect and contain a data breach in 2023 being 204
days~\cite{ibm2023databreach:online}} For example, the 2020 SolarWinds hack
persisted in the network for 14 months before being detected. The 2023 HHS
ransomware attack persisted in the network for 9 months before being detected.
The 2023 23andMe breach took place over a duration of 5 months before being
detected.




Enterprises too employ various systems to detect and respond to such attacks --
those that monitor endpoint or host activities (e.g., Endpoint Detection and
Response systems), those that monitor network traffic (e.g., Intrusion and
Detection Systems, or Intrusion Prevension Systems), monitor user activity
(e.g., User and Entity Behavior Analytics), aggregate and monitor system
information such as logs (e.g., Security Information and Event Management). These systems are often deployed on endpoint devices (e.g., servers,
laptops), inside network infrastructures (e.g., routers, switches), or in
centralized hardware (e.g., Active Directory server), and are passively
monitoring for threats. While a combination of these systems is deployed in
modern enterprises today, a team of analysis are tasked to triage and respond to
threats these systems generate. It is this step that is labor intensive and time consuming, and often
has the highest risk of missing a true positive (commonly termed the
\textit{needle-in-the-haystack} problem). As a result, several systems have been
proposed to aid the investigative process by tracking related events, piecing
together a causal reasoning of those events that led to the attack (termed
"causal analysis"), and understanding the scope of the attack, all in an effort
to minimize false positives, and accelerate the triage process.

Unfortunately, the scale and complexity of an enterprise poses several challenges in building such systems. First, the sheer volume of data that needs to be analyzed is large. Second, the data is often noisy, and is heavily skewed towards benign activities. Third, the data is often spread across multiple systems, and piecing together a causal graph is challenging. As a result these systems employ numerous heuristics and models to prevent false positives, and make the investigative process tractable. For example, systems often prune irrelevant dependencies, and only track those that are relevant to the attack, or prioritize alerts based on the severity of the attack, and the likelihood of the alert being a true positive. Systems also use a variety of techniques to reduce the number of alerts that need to be investigated, including using graph topological properties, or rareness scores to prioritize alerts.

The drawback however is that automated investigation systems assume an ideal world
-- one where the system has access to all the data, every event is logged, and
often ignore the practical challenges in analyzing such threats (e.g., the time
cycles an analyst spends on a threat, missing or restricted access to data). In this paper, we survey the
state-of-the-art in enterprise threat response systems. We first describe the
challenges in building such a detection or investigation system. We then describe
the different types of systems proposed by prior work, the heuristics or models they use,
and end with a discussion of open problems in this space.

\section{Background}

In this section, we start by discussing
Endpoint Detection and Response (EDR) tools, a suite of tools widely deployed within todays enterprises, and their use in hunting for threats. We then discuss the origins of automated attack investigation, followed by a birds-eye view of a typical attack kill chain. Lastly, we present a classification of systems that perform automated attack investigation.

\subsection{Classic EDR and SIEM Systems}

EDR systems or tools are pieces of software installed on endpoints (such as laptop, computers, mobile devices) within an enterprise, to continuously monitor low-level system activities (such as processes, network, files, etc). These work by rule-matching events to a known database of curated rules (such as MITRE ATT\&CK~\cite{mitreattack:online}) and flag threats that positively match. For matched events, the EDR tool can query additional context from the endpoint, and optionally stream the data to a centralized server for further analysis. Systems like IBM QRadar~\cite{qradar:online}, Splunk~\cite{splunk:online}, and ElasticSearch~\cite{elastic:online} are examples of centralized servers that collect and analyze data from EDR systems. These are often termed as Security Information and Event Management (SIEM) systems, and offer an interface for threat analysts to query, merge related events, and analyze the veracity of these warnings.

However, while these systems can provide good coverage in tracking low-level
system events, the burden often lies on the analyst to leverage their skill, and
expertise to piece together the context of an attack. Indeed, SIEM rules aid by
automatically detecting signatures or rules from collected events.
Unfortunately, these rules aren't bullet-proof and can often be
evaded by subtly modifying the attack to avoid detection via
rule-matching~\footnote{Indeed Uetz et al.~\cite{uetz2023you} showed in their
recent USENIX '23 paper that nearly half of the current SIEM rules could be
easily evaded by adversaries by using simple evasion tactics.}.

The industry's adoption of EDR systems has been sudden and quick, with nearly
56\% of organizations in 2021 having begun to deploy these within their
enterprise, according to a survey~\cite{comcastedr:online}. As a result,
enterprises today have contractual obligations with their EDR vendor (e.g.,
Qualys, FireEye), to deploy EDR tools in a low-cost, scalable manner within
their enterprise, plugged in and ready to capture threats\footnote{Often
motivated by indirect reasons, e.g., lowering the premiums on your cyber
insurance policy.}. This
further compounds the fact that there is ultimately a ``human-in-the-loop'' to
make sense of the various alert, and indeed. However, nearly 70\% of
organizations have reported that they struggle to keep up with the volume of
alerts these EDR systems generate~\cite{kasperskyesg:online}. In parallel, a
myriad of automated attack investigation systems have been proposed in recent
literature to aid the process of threat investigation. 

\begin{figure*}[]
    \centering
    \includegraphics[width=\textwidth]{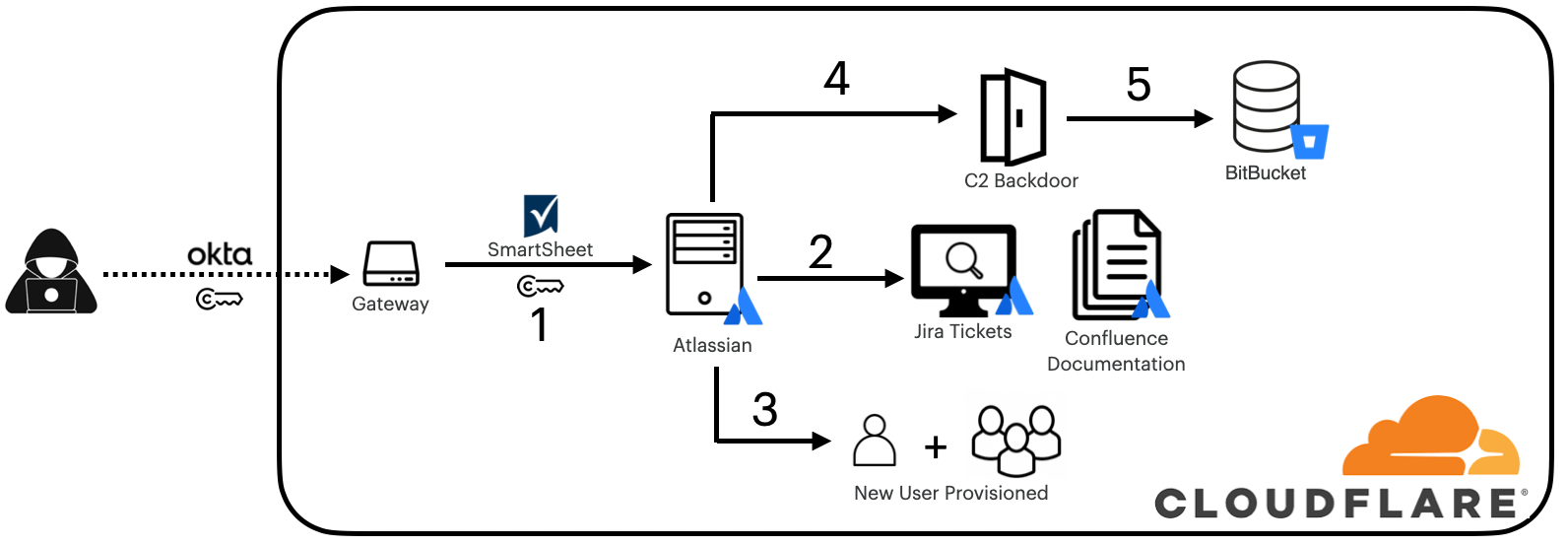}
    \caption{Reconstruction of an APT style attack kill chain based on a real-world incident at Cloudflare detected on November 2023. The sequential numbering represent the timeline of the attack as reported by Cloudflare's internal investigation~\cite{cloudflareincident:online}.}
    \label{fig:cloudflare-incident}
\end{figure*}














\subsection{Origins of Automated Investigation}


Perhaps the earliest work that performs automated attack investigation is BackTracker~\cite{king2003backtracking}. BackTracker uses system-call logs at the hypervisor level to piece together a graph that is used for investigating. The idea of using logs (as an auditing mechanisms) for attack investigation stems from ReVirt~\cite{dunlap2002revirt}. ReVirt replays system events in a Virtual Machine to analyze intrusions.  Similarly, systems like Taser~\cite{goel2005taser}, and Back-to-the-Future~\cite{hsu2006back} use logging to achieve similar goals of recovering system state. Taser uses kernel-based logging to do so, while Back-to-the-Future presents a framework for automatically rolling back system state to a previous, safe checkpoint, and only replaying the "clean" events. 

The idea of using causality graph to aid in investigation originates from King
et al.~\cite{king2005enriching}, who use a combination of network IDS (Snort)
and host IDS to verify the veracity of the alerts. Sitaraman et
al.~\cite{sitaraman2005forensic} also propose augmenting existing IDS by logging
additional parameters which is then used during analysis. Similarly, the use of
investigative tools for analyzing malware has also led to several proposed
systems~\cite{yin2007panorama, vasudevan2006cobra}.
Panorama~\cite{yin2007panorama} analyzes taint graphs (at a hardware-level) to
identify suspicious information access and processing behavior of foreign code.
Cobra~\cite{vasudevan2006cobra} presents a dynamic code analysis tool that
tracks fine-grained information flow to identify malicious code. 



\subsection{Modern Investigation Systems}

\begin{table*}[t]
    \centering
    \begin{tabular}{l||ccccc}
        \toprule
        \toprule \textbf{System} & \textbf{Attack Focus} & \textbf{Data} & \textbf{Host} & \textbf{Detection} & \textbf{Provenance}\\
        & \textbf{} & \textbf{Source} & \textbf{Coverage} & \textbf{Type} & \textbf{Type}\\
        \midrule
        Hopper~\cite{ho2021hopper} & Lateral Movement & Log-in & Multi & Offline & - \\
        \ltgrey Latte~\cite{liu2018latte} & Lateral Movement & Kerberos requests & Multi & Offline & - \\
        SAL~\cite{siadati2017detecting} & Lateral Movement & Log-in & Multi & Offline & - \\
        \ltgrey PrioTracker~\cite{liu2018towards} & APT & Provenance & Multi & Offline & Coarse \\
        RapSheet~\cite{hassan2020tactical} & APT & Provenance & Multi & Online & Coarse \\
        \ltgrey NoDoze~\cite{hassan2019nodoze} & APT & Provenance & Multi & Offline & Coarse \\
        ProvDetector~\cite{wang2020you} & Stealthy Malware & Provenance & Multi & Online & Coarse\\
        \ltgrey AirTag~\cite{ding2023airtag} & APT & Provenance & Multi & Offline & Coarse \\
        Morse~\cite{hossain2020combating} & APT & Provenance & Multi & Offline & Coarse \\
        \ltgrey SEAL~\cite{fei2021seal} & Causality Analysis & Provenance & Multi & Offline & Coarse \\
        NodLink~\cite{li2023NodLink} & APT & Provenance & Multi & Online & Coarse\\
        \toprule
    \end{tabular}
    \caption{Classifying 10 systems based on their capability, data source, hosts covered, system type, and detection type (APT -- Advanced Persistent Threat).}
    \label{tab:system-classification}
\end{table*}

Systems that perform automated attack investigation vary on several parameters
-- data source, investigation type, system type, etc. Perhaps the broadest
classification we can make is based on the detection system itself, i.e., if it
is online or offline system. Online systems are those that detect attacks in a
real-time (e..g, NodLink~\cite{li2023NodLink}), while offline systems work by
analyzing already collected logs or events (e.g., Airtag~\cite{ding2023airtag},
ProvDetector~\cite{wang2020you}, OmegaLog~\cite{hassan2020omegalog}). Based on
the data source used for investigation, systems can be classified as
provenance-based (e.g., Morse~\cite{hossain2020combating},
NoDoze~\cite{hassan2019nodoze}, PrioTracker~\cite{liu2018towards}), or other
data sources (e.g., Hopper~\cite{ho2021hopper}, Latte~\cite{liu2018latte}).
Fine-grained systems exploit provenance by tracking OS-level events, while
coarse-grained systems track specific components such as Active Directory log-on
events.

Systems can also be classified as single-host and multi-host based on if their investigation capacity. Single host systems (e.g., ProvDetector~\cite{wang2020you}) and multi-host systems (e.g., RapSheet~\cite{hassan2020tactical}). While single host systems track events within a single host, they fail to catch lateral movement, APT style attacks that traverse across multiple systems and networks. Lastly, provenance systems can further be classified as those that track fine-grained provenance and coarse-grained provenance. The former maps each output byte to its input byte, incurring high overheads in the process~\cite{xu2006taint, newsome2005dynamic}, while the latter tracks information flow at a coarser level~\cite{hossain2020combating, hossain2017sleuth}. Table~\ref{tab:system-classification} presents a classification of 10 systems we study in this paper based on the above taxonomy.

\subsection{Birds-Eye View of an Attack Kill Chain}

To constrast these tools and the nature of their design, we present a modern APT style attack kill chain.
Figure~\ref{fig:cloudflare-incident} presents a birds-eye view of an attack kill chain based on a real incident report at Cloudflare~\cite{cloudflareincident:online}. The kill chain starts with the attacker gaining access to three systems using unrotated credentials from the okta credential breach~\cite{cloudflareokta:online}. The attackers first authenticates with the Gateway using an access token they (step 1), and proceeds to access the Atlassian Server using a service account credential(associated with \textit{SmartSheet} SaaS application). The attacker then proceeds to do reconnaissance for a period of time by accessing Atlassian tools like Jira, and Confluence, and viewing support tickets and product documentation (step 2). The attacker then creates a new user and adds them to privileged groups, as a measure to ensure their persistence in the event the SmartSheet credential gets revoked (step 3). Further, to establish persistence, a Command-and-Control (C2) backdoor software (Sliver Adversary Emulation Framework) is installed and used to exfiltrate data. Lastly, the attackers accessed BitBucket Code repositories (again using compromised credentials), and download several repositories of data. This attack was detected by the Cloudflare team (nearly a month from initial compromise, by an alert) and the user accounts were deactivated, and the C2 backdoor was removed. The attackers also failed to laterally move but access a non-production console server in a different data center (based on a non-enforced ACL) but no evidence of further access was found. 










\section{Challenges}
\label{sec:challenges}




The classic systems used in enterprises suffer from a rates of high alert reporting/alarms (and high \textit{false positives}). Analysts are overwhelmed by the number of alerts, and is unable to respond to all of them -- a condition termed ``Alert Fatigue''. Compounding this problem is the fact that debugging each alert is a time-consuming, manual process of reverse-engineering the context, querying SIEM tools for more context, pulling logs from different endpoints EDR tools if needed, and so on. Missing a true positive when assessing the veracity of these alerts has a high cost. Alas, attacks in the wild such as the 2013 Target data breach~\cite{targetbreach:online} can be attributed to this very reason.

Hence, a fundamental challenge of automated attack systems is to reduce the number of false positives, while maintaining accurate detection rate. However, in doing so, a collection of new problems emerge. Dependencies explode as more dependencies are tracked to verify truthfulness around each alert. Storage size for logs and events increase exponentially, as more volume and variety of events are logged.\footnote{Fei et al.~\cite{fei2021seal} report that a typical enterprise produces 50GB amount of logs daily, from a group of 100 hosts.} Performance of the systems is time-critical and should not be a bottleneck in the detection process.

Even when constructed, such systems have a challenging problem evaluating their accuracy or performance. The problem lies in the fact that these systems rely on benign or normative data to train, and obtaining real-world benign enterprise data is challenging. Such data is often proprietary and reveals sensitive information about the structure and operations of the enterprise. Even when such data is available, the imbalance in real threats versus benign data makes it difficult to evaluate the performance of these systems (one event in a million may be malicious). Systems have resorted to using synthetic data, or simulating attacks to evaluate their performance, which may not be representative of real-world attacks. 

Despite these challenges, the reality is that a ``human-in-the-loop'' approach is still the norm in many enterprises. The analyst has a limited time-budget (e.g., 15 minutes per alert) to investigate and respond to each alert. Therefore, systems should to be able to prioritize ``more important'' alerts, and provide the analyst with interpretable information to make a timely decision.












\section{Attack Investigation Approach}

The first step in attack investigating involves \textit{causality analysis}, that is, inferring the root cause of the attack and its scope by linking dependent events. This is done typically by building a causal graph modeling events or logs that are collected within the enterprise. The causal graph can be constructed using different types of data (e.g., Hopper~\cite{ho2021hopper} and Structurally Anomalous Logins (SAL)~\cite{siadati2017detecting} detector use login information from system logs, Latte~\cite{liu2018latte} uses Kerberos Service Ticket Requests, Morse~\cite{hossain2020combating}, NoDoze~\cite{hassan2019nodoze} and other provenance-tracking systems exploit low-level OS related events like files and processes). The second step in attack investigation is prioritize anomalous events or paths (e.g., PrioTracker~\cite{liu2018towards} uses graph topological properties and rareness score to prioritize anomalous paths, NodLink~\cite{li2023NodLink} also uses topological features modeled as a tree). Simultaneous, systems tend to deprioritize paths based on other heuristics (e.g., PrioTracker~\cite{liu2018towards} deprioritizes paths without high fanout).

\begin{figure*}
    \centering
    \includegraphics[width=\textwidth]{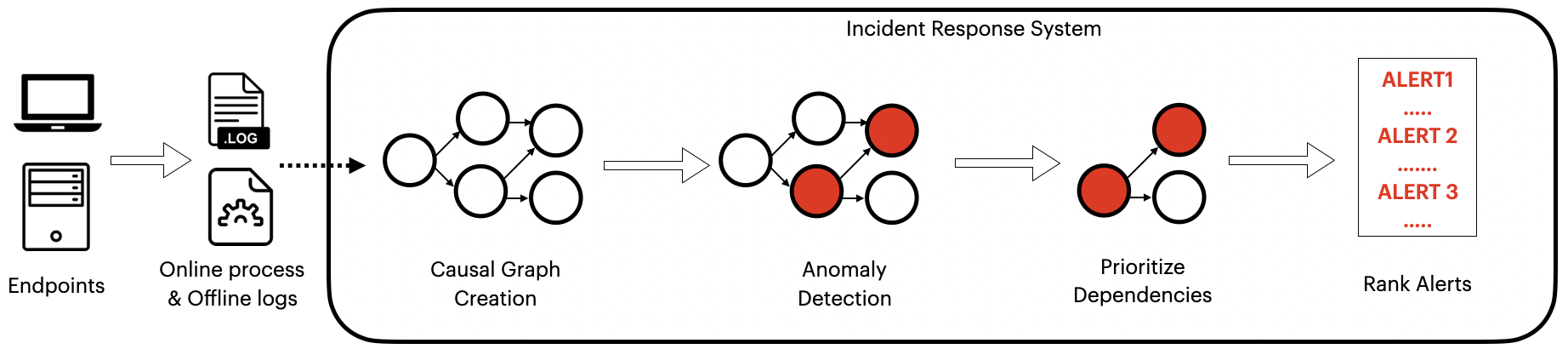}
    \caption{General architecture of systems that perform automated attack investigation.}
    \label{fig:systems}
\end{figure*}

\subsection{Causality Graph Construction}

In order to establish the veracity of a warning, the system performs causality analysis to determine the root cause (aka root cause analysis). When building a causal graph, the system exploits data provenance, logs, or other data sources it is tracking. Below we discuss three different systems that exploit different data sources to build a causal graph -- Hopper, Latte, and PrioTracker. Hopper~\cite{ho2021hopper} uses login information from system logs to build a login graph. Each edge forms a login event from a source machine to a destination machine, timestamped at the time of the login. Hopper then infers a causal user, a user who originally logged in, resulting in this edge, along with the a set of rareness rules to decide if the event is an example of lateral movement. 
Latte~\cite{liu2018latte} instead uses Kerberos Service Ticket Request events (filtered based on a specific windows event alert code) to model a graph of computer-to-computer user logins. Similar to Hopper, each edge is directed from a source node to a destination node, timestamped at when the connection was made. Latte then uses a set of heuristics to determine if the event is an example of lateral movement. PrioTracker~\cite{liu2018towards} like other provenance systems, builds a dependency graph of OS-level system events. Each node in the graph represents a system object (e.g., process, network socket, or file), and the edges represent the flow of data (e.g., process creation, file read, etc.). Again, multiple edges are temporally ordered. PrioTracker then uses a set of heuristics to associate priority with every causal event that it uses during investigation.

\subsection{Root Cause and Scope Analysis}


Once the causal graph has been constructed, the system determines the root cause and the scope, using forward and backward tracing respectively. The root cause is inferred using backward tracking, starting from a known malicious event and tracing incident edges to determine the source. The scope, in contrast, is determined using forward tracking, starting from the known malicious event, and tracing forward by following incident edges. During both these cases, the system causally tracks an immense number of paths that explode exponentially based on linked dependencies. In section~\ref{sec:solutions} we'll discuss how systems mitigate this dependency explosion during their analysis. For now, we discuss four systems from prior literature, RapSheet, NoDoze, PrioTracker, and Hopper that perform root cause and scope analysis.

RapSheet~\cite{hassan2020tactical} uses Symantec's EDR system to build a~\textit{Tactical Provenance Graph} by tracking Initial Infection Points (IIP) vertices from the provenance graph. By defining a set of rules that involves backtracking from a known malicious event, RapSheet determines the IIP vertices, and then issues forward traces from these points to generate IIP graphs that it uses to model the scope of the attack. NoDoze~\cite{hassan2019nodoze} directly uses progeny and ancestry of the infected node to determine if the alert is true attack or a false alarm, using backward and forward propagation. PrioTracker~\cite{liu2018towards} introduces the notion of priority when performing forward and backward tracking. The system uses a set of heuristics to score each path with a priority score, and picks the path with the highest priority score during forward and backward propagation. Hopper~\cite{ho2021hopper} produces a sequence of causally-related logins using backward-tracing. This is based on a heuristic that two edges are correlated if they are incident on the same vertex, and are within a threshold time window. Hopper then uses a set of rules to determine if the path is an example of lateral movement. 




\subsection{Anomaly Detection}
\label{sec:anomaly-detection}

During root cause and scope analysis, systems employ various heuristics to detect anomalous paths. These heuristics can be based on topological graph features, machine learning models, or other relationships, and are used to score paths based on their likelihood of being malicious.

\subsubsection{Topological Graph Heuristics}
\label{sec:topological-graph-heuristics}

For those systems that use topological graph features, the scoring is based on event granularity, the whole path, or parts of the whole graph. We discuss each of these three types using PrioTracker, NoDoze, Hopper, and RapSheet as examples.

PrioTracker~\cite{liu2018towards} assigns a priority score (a measure of suspiciousness) to each event representing an edge in the causal graph, using a combination of rareness and fanout score. While fanout can be calculated using the number of incident edges at a node, rareness is a function of deviation from normal behavior. PrioTracker thus uses a reference model of system events (collected from 150 hosts within an enterprise) to define ``normal behavior'', and measures rareness as a deviation.

NoDoze~\cite{hassan2019nodoze} in contrast assigns an anomaly score to the entire path by associating the contexts in which each event occurs. This helps NoDoze distinguishing between legitimate use of a software (e.g., cclearner for cleaning disk) and its malicious use (ccleaner used by ransomware for destroying evidence). To do so, NoDoze uses a network diffusion algorithm to combine scores from events along a path. NoDoze generates a regularity score to each dependency path, which is a combination of fanout (using IN and OUT scores), and rareness (using historical window). NoDoze then normalizes the anomaly score using a decay factor that prevents the anomaly scores of longer paths from being skewed.


Hopper~\cite{ho2021hopper} also uses a path-based scoring heuristic modeled on credential switches -- if a path involves a credential switch to access a resource that the user has not accessed recently (30 days), then it is considered malicious. If the credential switch in the path is less clear, Hopper uses a probability scoring model to score events along a path based on their historical context.

RapSheet~\cite{hassan2020tactical} relies on constructing a subgraph (a \textit{Tactical Provenance Graph (TPG)}) by identifying the initial points of compromise in a causal graph. Such graph-based systems capture scenarios where an attacker spawns multiple processes after compromise, while a path-based system treats each of those causally related paths different.  RapSheet relies on MITRE's ATT\&CK framework~\footnote{A curated set of expertly-written rules based on analysis of real-world APT attacks~\cite{mitreattack:online}.} to categorically match and score the graph, maximizing the threat score of the TPG. For each MITRE techniques that is matched, RapSheet uses a categorical score for severity and likelihood (e.g., 1 for very low, 5 for very high, etc.) which it weights and combines to produce a final score for the TPG. Intuitively, the TPG graph is a combination of the sequences of possible kill-chain of the APT attack, temporally ordered, and scored using the ATT\&CK framework.


\subsubsection{Machine Learning Models}

A subset of ML models use Natural Language processing (NLP) methods to directly process logs, a task well suited for unstructured text~\cite{ding2023airtag, wang2020you}. Other systems use ML techniques to capture information from provenance graphs and learn  benign vs malicious behaviors automatically. We discuss ProvDetector, Airtag, and NodLink as examples.


ProvDetector~\cite{wang2020you} tracks events at the OS-level tracker to build a per-process provenance graph aimed at detecting stealthy malware. It then finds the K-rare paths in the graph (using an idea based on regularity scoring similar to NoDoze~\cite{hassan2019nodoze}) and uses doc2vec to learn the embeddings of these paths. This method has an advantage that it is self-supervised, since the model can learn based on purely benign data. Lastly, it uses a classification based on threshold of embedded vectors detected as anomalous, to determine if the process is malicious.

AirTag's~\cite{ding2023airtag} main intuition is that logs and causal graphs are different representations of the same problem, however logs are in Euclidean space, and easier for ML models to learn. Thus, AirTag works by directly training on benign logs (preprocessed and embedded with a customized BERT language model~\cite{devlin2018bert}) and learns a decision boundary using one-class Support Vector Machine (OC-SVM~\cite{sklearnoc:online}). It then uses the same embedding during testing to detect anomalous log entries that lie outside the decision boundary. However, AirTag reconstructs the detected attack causal graph from the anomalous log entries to present the result to the analyst.

NodLink~\cite{li2023NodLink} constructs a Steinier Tree from the provenance data by searching for a tree that minimizes the number of edges. The tree spans anomalous events and IOCs with minimum edges, and has a theoretical guarantee of finding such a subgraph in polynomial time. However, to identify suspicious processes in an online manner, NodLink uses a document embedding technique to embed processes, and uses a Variational Auto Encoder (VAE) to generate an anomaly score. The anomaly score is then used to determine if the process is malicious.

\section{Solutions}
\label{sec:solutions}

Each system discussed so far (Section~\ref{sec:challenges}) addresses the challenges of classic EDR systems (e.g., alert fatigue, storage limitations, etc.) in their individual, creative ways. The common themes involve mitigating dependency explosion, prioritizing alerts, mitigating storage requirements, and improving interpretability. Table~\ref{tab:soln-comparison} scores each of the systems based on these capabilities.  In this section, we discuss a subset of systems when discussing each capability in detail.



\begin{table*}[t]
    \centering
    \begin{tabular}{l||ccccc}
        \toprule
        \toprule \textbf{System} & \textbf{Mitigating Dependency } & \textbf{Prioritizing Alerts} & \textbf{Interpretable } & \textbf{Mitigating Storage} \\
        & \textbf{Explosion} & & \textbf{Results} & \\
        \midrule
        Hopper~\cite{ho2021hopper} & Deduplicate paths, & Time-budget & \xmark & \xmark &\\
        & optimized backtracing & & & &\\

        \ltgrey Latte~\cite{liu2018latte} & \xmark & Path ranking & \xmark & Graph pruning\\

        SAL~\cite{siadati2017detecting} & \xmark & \xmark & \xmark & \xmark \\

        \ltgrey PrioTracker~\cite{liu2018towards} & Prioritize dependencies & Anomaly scoring & \xmark & \xmark \\

        RapSheet~\cite{hassan2020tactical} & Concise graph & \xmark & Concise graphs & Graph pruning  \\

        \ltgrey NoDoze~\cite{hassan2019nodoze} & Execution partitioning & Anomaly scoring & \xmark  & \xmark \\

        ProvDetector~\cite{wang2020you} & Path sampling  & \xmark & \xmark & \xmark \\

        \ltgrey AirTag~\cite{ding2023airtag} & -  &  \xmark & Concise graphs  & \xmark \\

        Morse~\cite{hossain2020combating} & Tag propagation  & \xmark & \xmark  &  \xmark \\

        \ltgrey SEAL~\cite{fei2021seal} & \xmark & - & \xmark &  Graph compression \\

        NodLink~\cite{li2023NodLink} & Threshold cutoff & Anomaly scoring & \xmark &  In-memory cache  \\

        \toprule
    \end{tabular}
    \caption{Scoring 11 systems based on their capability to solve challenges of classic EDR tools. \xmark  indicates the system does not have that capability, and - indicates the capability is not applicable.}
    \label{tab:soln-comparison}
\end{table*}

\subsection{Mitigating Dependency Explosion}

With an increase in provenance tracking, systems gather an abundant amount of fine grained data spanning multiple hosts, and causally links these events with their predecessors in time. For long running processes that involve large data flows, the number of dependencies can grow exponentially (e.g., a long running browser process). Systems therefore have used ingenious ways to mitigate this dependency explosion. We discuss traditional systems (e.g., ProTracer) and recent systems (e.g., Morse, NoDoze) that  address this challenge.

The traditional approach to handle dependency explosion has been to partition long running processes into individual units to which dependencies are confined (e.g., an output can only be dependent on an input in the same unit/partition). Indeed, systems like ProTracer~\cite{ma2016protracer}, BEEP~\cite{lee2013high}, MCI~\cite{kwon2018mci}, and MPI~\cite{ma2017mpi} use this approach to provide a more granular view of the system. Unfortunately though, these systems require user interactions, kernel modifications, or are limited to specific operating systems (e.g., Linux). 

Morse~\cite{hossain2020combating} associates tags to subjects and objects within the system that it attenuates and decays as data propagates through the system. Specifically Morse assigns two types of tags -- subject tags that it uses to determine if the process or the environment is benign or suspicious, and data tags that captures the confidentiality and integrity of values. It then uses tag propagation rules to determine the value of the tags as they propagate through the system. For example, tags for benign subjects are attenuated, and decayed periodically with a quiescent factor. This way, the system can prune irrelevant system dependencies, and focus on the relevant ones.

NoDoze~\cite{hassan2019nodoze} specifically handles long running processes by using a partitioning technique (termed \textit{behavioral execution partitioning}) that separates the normal and anomalous behavior of long running processes. NoDoze does this by generating a \textit{True Alert Dependency Graph} that includes only relevant, true dependencies of particular length ($\tau_l$), with a anomaly score higher than a threshold ($\tau_d$). To generate the anomaly score, NoDoze uses topological features such as regularity score, and fanout (see Section~\ref{sec:anomaly-detection}), and merges anomalous paths to reduce the number of alerts generated.

ProvDetector~\cite{wang2020you} uses a path sampling technique to mitigate dependency explosion. From each per-process dependency graph, ProvDetector samples a set of paths that are representative of the graph. It then uses these sampled paths to generate embedding and classify the process as benign or malicious.

\subsection{Prioritizing Alerts}

While mitigating dependency explosion when backtracking an attack is important, it need not reduce the number of alerts generated, and hence may not reduce the time budget for an analyst. We discuss three systems that associate priority and rank alerts based on their importance -- Hopper, and PrioTracker.

Priotracker~\cite{liu2018towards} uses a priority assignment algorithm that is a combination of rareness and fanout scores to assign priority to alerts (Section~\ref{sec:topological-graph-heuristics}). Both the rareness and fanout scores are weighted in a combined priority assignment using weights learnt from events within an enterprise. In practice, PrioTracker uses a threshold for rareness to assign alarms to categories of alerts (e.g., Critical events, Rare critical events).

Hopper~\cite{ho2021hopper} makes a distinction with login paths with clear credential switch (which is earlier to analyze) and unclear credential switch (based on a probabilistically heuristic). Hopper uses a user-inputted time budget when generating alerts for paths with unclear credential switches. This way, Hopper can prioritize alerts based on the time budget of the analyst. For each unclear path P, Hopper extracts three features (based on path characteristics) and computes an anomaly score. Then, the anomaly score is ranked relative to a historical paths and if within a window of 30 $\times$ alert-budget, Hopper generates an alert for the path.


\subsection{Mitigating Storage}

System logs within an enterprise can grow exponentially over time as more dependencies get tracked. Throwing out old logs with sensitive information can hinder the ability to backtrack an attack and decipher its root cause. Indeed, to detect long running APT attack campaigns, the ability to retain old logs becomes crucial. To avoid this, the general approach taken relies on pruning graph sizes and hence compressing the associated log storage to save storage space ~\cite{fei2021seal,hassan2020tactical}. We discuss RapSheet, and SEAL as examples.

RapSheet~\cite{hassan2020tactical} uses a graph-based pruning approach to compact storage by applying two graph reduction rules.  First, RapSheet removes object nodes if they have no alert dependencies in the backward chain of the TPG, and no alert edges incident. Second, RapSheet removes process nodes that have no alert events in the backward chain of the TPG, and if the process has already been terminated. Applying these two rules, RapSheet periodically runs the graph reduction algorithm to generate a space-efficient TPG that can be stored for long periods of time.

Unlike prior work that does lossy compression (e.g., ~\cite{hossain2018dependence, tang2018nodemerge}), SEAL~\cite{fei2021seal} uses a lossless compression technique that ensures it can be queried without much decompression overhead by an analyst. To do this, SEAL ensures \textit{decompression-free} compression on graph structure by merging edges with similar patterns, and \textit{query-able} compression on edge properties by using different coding (delta, golomb) to compress edge properties (like timestamps). Further, SEAL selectively decompresses only necessary parts of the graph by comparing timestamps of the compressed edge and that of the query.

\subsection{Interpretable Results}

While both mitigating dependency explosion and prioritizing alert help the automated triaging of alerts, ultimately, the issue is flagged to a human analyst for further analyst. Thus, making the result interpretable and providing sufficient context to validate the threat is crucial. We discuss how AirTag, and Dong et al.~\cite{dong2023we} address this challenge.

Having prioritized logs with anomalous behaviors, AirTag~\cite{ding2023airtag} generates anomalous causal graphs for the analyst to investigate. AirTag uses a greedy algorithm by generating disjoin graphs for every logs flagged by the classifier (OC-SVM) and merging them by finding the first common node between disjoin graphs. AirTag only tries to find the first common node between disjoin paths, and is able to recover the attack story (based on its evaluations) by repeating this approach.

Dong et al.~\cite{dong2023we} conduct a survey on industrial viewpoint of different provenance system,  Indeed, most industry managers view interpretability as a crucial feature which adds more context and aids in investigation, and that basic provenance systems that consists of low-level audit events from logs are understandable for skilled analysts, as long as they are concise. However, when reimplementing different provenance system from prior work, Dong et al. found that atleast one system (Unicorn~\cite{han2020unicorn}) has a low alert frequency (alarms/host/day), but reports the whole provenance graph as an attack alarm, without pinpointing the location within. Such tradeoffs in interpretability can hinder the ability of an analyst to validate the threat. We defer a more detailed discussion of this to Section~\ref{sec:tradeoffs}.





\section{Open Problems}
\label{sec:tradeoffs}

\begin{table*}[]
    \centering
    \begin{tabular}{l||l|l}
        \toprule
        \toprule \textbf{System} & \textbf{Benign Data Source} & \textbf{Attack Data Source} \\
        \midrule
        Hopper~\cite{ho2021hopper} & Enterprise data (Dropbox) & Red-team testing, \\
        & & Simulated attacks \\

        \ltgrey Latte~\cite{liu2018latte} & Enterprise data (Microsoft and & 
        Red-team testing, \\
        \ltgrey & unnamed organization) & Real-world attack \\

        SAL~\cite{siadati2017detecting} & Enterprise data (Unnamed financial company) & Simulated attacks \\

        \ltgrey PrioTracker~\cite{liu2018towards} & Enterprise data (Unnamed IT organization) & Simulated attacks \\

        RapSheet~\cite{hassan2020tactical} & Enterprise data (Symantec) &  Red-team testing, \\
        & & Simulated attacks  \\

        \ltgrey NoDoze~\cite{hassan2019nodoze} & Enterprise data (NEC Labs America) & Simulated attacks \\

        ProvDetector~\cite{wang2020you} & Enterprise data (Unnamed organization) & Simulated attacks \\

        \ltgrey AirTag~\cite{ding2023airtag} & Public datasets (ATLAS, DEPIMPACT),  & Simulated attacks,  \\
        \ltgrey & & Public datasets (ATLAS, DEPIMPACT)\\

        Morse~\cite{hossain2020combating} & Public datasets (DARPA TC) &  Public datasets (DARPA TC)\\

        \ltgrey SEAL~\cite{fei2021seal} & Enterprise data (Unnamed enterprise), & Simulated attacks, \\ 
        \ltgrey & Public datasets (DARPA TC) &  Public datasets (DARPA TC) \\

        NodLink~\cite{li2023NodLink} & Public datasets (DARPA TC datasets), 
         & Public datasets (DARPA TC datasets), \\ 
        & Enterprise data (Sangford) & Red-team testing, \\
        & & Real-world deployment (Sangford) \\

        \toprule
    \end{tabular}
    \caption{Evaluation datasets used by 11 systems in prior literature.}
    \label{tab:eval-comparison}
\end{table*} 

Section~\ref{sec:solutions} discusses creative way of addressing new challenges in attack investigation, namely mitigating dependency explosion, managing storage, prioritizing alerts, and presenting interpretable results. However, new challenges arise as a result of these solutions. We discuss these challenges in this section, and propose a set of open problems that future work should consider.

\subsection{Push towards Real-time Analysis}

8/11 systems we study (Table~\ref{tab:system-classification}) are designed for offline analysis. The motivation is that offline analysis allows more comprehensive analysis, and does not need to be resource constrained as online analysis. However, offline analysis has its own limitations. For instance, Verizon Data Breach Report~\cite{verizondatabreach:online} found that it takes a matter of seconds to compromise a system, and then several minutes to hours for exfiltration. This means that offline analysis may not best geared to detect breaches in real-time. Indeed, systems like ProvDetector~\cite{wang2020you} and NodLink~\cite{li2023NodLink} propose an online detection algorithm, and we believe that this is a promising direction for future work.

We also believe that the bottleneck for online systems does not lie in the storage. Indeed, a number of creative systems have proposed compressing the provenance graph (lossless and lossy)~\cite{hossain2018dependence, tang2018nodemerge, fei2021seal}, pruning the size of the graph~\cite{}. Instead, we believe a combination of storage factored with performance is more important. For instance, NodLink~\cite{li2023NodLink} prunes the dependencies with a threshold, and then uses a combination of in-memory cache and disk storage to improve the performance. Other systems should also consider this tradeoff as a motivation for future work.

\subsection{Failure Reducing Analyst Workload}

Unfortunately, the real motivation behind automated analysis is to reduce the workload of the analyst. This is where a bulk of the systems we studied fall short. Indeed, 5/11 (Table~\ref{tab:soln-comparison}) systems prioritize alerts the system generates, and nearly all the systems we studied attempt to improve the quality of these alerts (e.g., by exploiting context around the event). However, only 1/11 systems consider the time cycles of the analyst as a parameter. Further, only 1/11 systems considers how interpretable the result is to the analyst. This is a significant limitation, and we believe that future work should consider the workload of the analyst as a parameter.

Dong et al.~\cite{dong2023we} also mentions that none of the systems in their evaluation accounted for the number of alerts/host/day raised, in an attempt to minimize it. Further, when reimplementing three systems from prior literature and evaluating their accuracy, they found only 1/3 systems (Unicorn~\cite{han2020unicorn}) with a low enough rate of alerts/host/day to be practical for real-world use. Ironically though, the same system had poor interpretability since it would flag the entire provenance graph as malicious to the analyst.

\subsection{Unrealistic Assumptions}

A number of systems make unrealistic assumptions in their design or evaluation. For instance, nearly all the systems we studied assume that the logs are tamper evident. While, this maybe a reasonable assumption, and systems indeed have been proposed to make logs tamper proof~\cite{bellare1997forward}, we believe systems should incorporate this as a feature in their design. Further, nearly all the systems we studied assume that the data is complete -- e.g., all the logs are available, all the log-in events are logged, access to all hosts or endpoints is unfettered. This indeed is a significant limitation, and in our own experience, we have found this to not be the case. Some systems are more well monitored that the rest, crucial logs might be missing, access to logs on some endpoint might be restricted, and so on. We believe that future work should consider these limitations in their design.

Besides these general observations, we also note a number of minor assumptions in the design of the systems we studied. For instance, Latte~\cite{liu2018latte} prunes systems with a high in-degree or out-degree, potentially allowing an attacker who purely hides in ``more popular'' systems to go unnoticed. SAL~\cite{siadati2017detecting} and Hopper~\cite{ho2021hopper} exploit the structure of the enterprise to detect attacks, which though is a reasonable assumption, might not be generalizable to all enterprises (e.g., SAL fails to consider if enterprises have no functional units, or if the functional units are not well defined).

Lastly, only one system we studied (RapSheet~\cite{hassan2020tactical}) considers a less invasive approach by building on top of existing EDR systems, while others use classic auditing systems like Auditd, or Sysdig. While these approaches aren't necessarily bad, we believe that future work should consider the adaptability of research solution in the real world enterprises, by using existing EDR systems as a base.

\subsection{Realistic Evaluations are Tough}

Alas the one challenge that remains the most difficult to address involves finding a high-quality dataset that is representative of real-world attacks. Table~\ref{tab:eval-comparison} shows the datasets used by the systems we studied for benign and attack data.

To generate benign data, systems partner with enterprises (e.g., Hopper~\cite{ho2021hopper} uses data from Dropbox, and NodLink~\cite{li2023NodLink} uses data from Sangford), or use public data traces with simulated attacks (e.g., Darpa TC datasets~\cite{darpatc:online}, ATLAS~\cite{alsaheel2021atlas}, DEPIMPACT~\cite{fang2022back}).  Finding high-quality dataset with real-world attacks on the other hand is a significantly more challenge -- both in terms of privacy, and in terms of enterprises being willing to share such data publicly. Hence, nearly all the systems we studied use simulated attacks, or red-team testing, or use simulated attacks from public datasets.

How realistic these simulations are to real-world attacks or representative of them is an open question. To the best of our knowledge only one system (Latte~\cite{liu2018latte}) uses real-world instance of Lateral Movement attack in their dataset for evaluation. 

Finally, even when a system proposed performs well on static traces, or synthetic data, it is not clear how well it would perform in the real-world. Again, to the best of our knowledge, only one system (NodLink~\cite{li2023NodLink}) reimplemented its system in an online setup to capture real-world threats. 

Thus, the quality of evaluation data, and its representativeness are both open questions that we urge future work to consider.

\section{Conclusion}

In this paper, we survey 11 state-of-the-art systems for automated attack investigation in enterprises. We find that these systems are designed to address challenges in attack investigation, namely dependency explosion, limited storage, alert fatigue, and do so with their own flavors.  However, there is no universal solution addressing each of these challenges, and systems therefore strike a balance in prioritizing some features more than the rest. We scored each of the 11 systems based on their techniques for overcoming these challenges, and extracted a set of similar approaches. We further discussed four open challenges in this line of work -- namely, the push towards real-time analysis, focus on reducing analyst workload, minimizing unrealistic assumptions, incorporating realistic evaluations. We believe that future work should consider these challenges in their design, and evaluation.

\begin{acks}
    Much thanks to Grant Ho, Stefan Savage and Geoffrey M. Voelker for their insights on this work.

\end{acks}

\bibliographystyle{ACM-Reference-Format}
\balance
\bibliography{reference}

\end{document}